\documentclass[10pt,conference]{IEEEtran}

\usepackage[utf8]{inputenc}
\usepackage[T1]{fontenc}    
\usepackage{hyperref}       
\usepackage{url}            
\usepackage{booktabs}       
\usepackage{amsfonts}       
\usepackage{nicefrac}       
\usepackage{microtype}      
\usepackage{enumitem}

\usepackage{bm}
\usepackage{graphicx}
\usepackage{doi}
\usepackage{amsfonts}
\usepackage{xcolor}
\usepackage{physics}
\usepackage{makecell}
\usepackage{multirow}
\usepackage{graphicx}
\usepackage{subcaption}
\usepackage{enumitem}
\usepackage{duckuments}

\usepackage{physics}
\usepackage{algorithm}
\usepackage{algpseudocode}

\usepackage{mathtools} 
\usepackage{array}
\usepackage{wrapfig} 
\usepackage{svg}
\usepackage[capitalise]{cleveref}
\usepackage[natbib=true, sorting=none, maxcitenames=1, style=numeric-comp]{biblatex}

\usepackage[colorinlistoftodos,prependcaption,textsize=tiny]{todonotes}

\usepackage{subfiles}

\title{Enhancing the Scalability of Classical Surrogates for Real-World Quantum Machine Learning Applications}
\author{Philip Anton Hernicht\IEEEauthorrefmark{1}\IEEEauthorrefmark{3}, Alona Sakhnenko\IEEEauthorrefmark{2}\IEEEauthorrefmark{4}, Corey O'Meara\IEEEauthorrefmark{1}, Giorgio Cortiana\IEEEauthorrefmark{1}, Jeanette Miriam Lorenz\IEEEauthorrefmark{2}\IEEEauthorrefmark{3} \\
\IEEEauthorblockA{\IEEEauthorrefmark{1}E.ON Digital Technology GmbH,  Munich, Germany}
	\IEEEauthorblockA{\IEEEauthorrefmark{2}Fraunhofer Institute for Cognitive Systems IKS,  Munich, Germany}
	\IEEEauthorblockA{\IEEEauthorrefmark{3}Ludwig-Maximilian University, Munich, Germany}
        \IEEEauthorblockA{\IEEEauthorrefmark{4}Technical University of Munich, Munich, Germany}
}
\date{\today}

\usepackage{amsthm}
\theoremstyle{definition}
\newtheorem{definition}{Definition}[section]

\begin{document}

\maketitle
\begin{abstract}
Quantum machine learning (QML) presents potential for early industrial adoption, yet limited access to quantum hardware remains a significant bottleneck for deployment of QML solutions. This work explores the use of classical surrogates to bypass this restriction, which is a technique that allows to build a lightweight classical representation of a (trained) quantum model, enabling to perform inference on entirely classical devices. We reveal prohibiting high computational demand associated with previously proposed methods for generating classical surrogates from quantum models, and propose an alternative pipeline enabling generation of classical surrogates at a larger scale than was previously possible. 
Previous methods required at least a high-performance computing (HPC) system for quantum models of below industrial scale (ca. 20 qubits), which raises questions about its practicality. We greatly minimize the redundancies of the previous approach, utilizing only a minute fraction of the resources previously needed. We demonstrate the effectiveness of our method on a real-world energy demand forecasting problem, conducting rigorous testing of performance and computation demand in both simulations and on quantum hardware. Our results indicate that our method achieves high accuracy on the testing dataset while its computational resource requirements scale linearly rather than exponentially. 
This work presents a lightweight approach to transform quantum solutions into classically deployable versions, facilitating faster integration of quantum technology in industrial settings. Furthermore, it can serve as a powerful research tool in search practical quantum advantage in an empirical setup.
\end{abstract}

\begin{IEEEkeywords}
Quantum machine learning, classical surrogates, energy demand forecasting, computational demand, quantum technology integration
\end{IEEEkeywords}

\section{Introduction}
Quantum machine learning (QML) represents a promising avenue for early adoption of quantum computing~(QC) algorithms in industrial use-cases. Significant progress has been made in the field, though the search for a "killer application" is still ongoing. A variety of QML algorithms have been proposed that demonstrate promising results compared to various classical benchmarks, e.g. \cite{Abbas_2021, Liu_2021,  rudolph2022generationhighresolutionhandwrittendigits, Glick_2024, agliardi2024mitigatingexponentialconcentrationcovariant}. 
However, once the breakthrough application is finally uncovered, a significant bottleneck remains that hinders the deployment of quantum solutions in industrial environments: limited on-demand access to quantum hardware. This significantly complicates the practical application of these algorithms, especially in real-time applications or in safety-critical areas where cloud access to QC hardware is not possible due to latency, and security requirements and regulations.

A popular choice of architecture for quantum circuits can be represented by a truncated Fourier series \cite{Schuld_2021}. This fact allows quantum models to be represented completely classically through their classical surrogates \cite{Schreiber_2023, landman2022classicallyapproximatingvariationalquantum}. This implies that once quantum models have been trained on quantum hardware, one can create a fully classical lightweight representation of their input-output mapping and deploy it in production even on edge devices. This classical representation makes it an attractive candidate for industrial adaptation as well as a testbed for practical quantum advantage.

In this work, we highlight the prohibiting computational demand associated with generating a classical surrogate as proposed in \cite{Schreiber_2023, landman2022classicallyapproximatingvariationalquantum}, which substantially restricts the size of classically representable quantum models well below industrial utility. The contributions of this work are three-fold: 
\begin{enumerate}[label=\Roman*.]
    \item We propose an alternative pipeline to create surrogates that significantly reduces the computational demands, enabling the conversion of substantially larger quantum models that was previously possible;
    \item We showcase the utility of this method by applying the pipeline to convert quantum models trained to perform an energy demand forecasting in a power plant of E.ON, and conduct an extensive investigation of required computational resources based on the solution requirements;
    \item We field-test our approach with Qiskit simulators and on an IBM quantum hardware by creating a classical surrogate of a quantum model that significantly surpasses the scale achievable with a simple device equipped with just 16 GB of RAM.
\end{enumerate}

This paper is structured as follows: We discuss previously proposed methods to create surrogates in \cref{sec:background}. We highlight the shortcomings of prior methods and propose improvements in \cref{sec:surrogation_2.0}. We empirically validate the effectiveness of our method through simulations and quantum hardware experiments, and we describe the experimental setup in \cref{sec:method}. We demonstrate the accuracy of our proof-of-concept implementation and provide a resource estimation in \cref{sec:results}. Finally, we discuss the implication of the existence of classical surrogates on possibility of quantum advantage with variational quantum models as well as future prospects in \cref{sec:discussion}.



\section{Background}\label{sec:background}
A classical surrogate of a quantum model needs to be lightweight, easy to generate and precise. Below, we outline the components of classical surrogates that ensure that these requirements are met.

\subsection{Variational quantum circuits as Fourier series}\label{sec:fourier-based}

\begin{figure}
    \centering
    \includegraphics[width=0.5\textwidth]{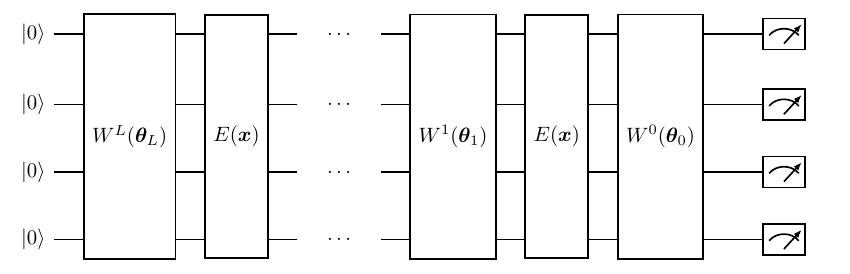}
    \caption{Variational reuploading model architecture~\cite{Schuld_2021} with $L$ layers, where $W$ represents learnable blocks with parameters $\bm{\theta}_{i \in [0, L]}$ and $E$ is an embedding block with an input vector $\bm{x}$.}
    \label{fig:reuploading}
\end{figure}

The variational reuploading quantum model $f_{\bm{\Theta}}(\bm{x})$ (illustrated in \cref{fig:reuploading}) is a prominent model type in the field of QML~\cite{P_rez_Salinas_2020, Jerbi_2023, Sakhnenko_2022, monnet2024understandingeffectsdataencoding}. \citet{Schuld_2021} showed that $f_{\bm{\Theta}}(\bm{x})$ can naturally be represented with a truncated Fourier series, which opens exciting avenues for analysis~\cite{Sakhnenko_2022, monnet2024understandingeffectsdataencoding, Barthe2024gradientsfrequency}. More importantly, it facilitates the classical representation of an input-output relationship of $f_{\bm{\Theta}}(\bm{x})$, which is the basis for the work presented in this paper. To make it more concrete, the models $f_{\bm{\Theta}}(\bm{x})$ are defined as follows:

\begin{equation}
    f_{\bm{\Theta}}(\bm{x}) = \expval{U(\bm{x};\bm{\Theta})^{\dagger} O U(\bm{x};\bm{\Theta})}{0},
\end{equation}
where $\bm{x}$ is the input vector, $\bm{\Theta}$ is a set of learnable parameters, $U(\bm{x};\bm{\Theta}) = W^L(\bm{\theta}_L) E(\bm{x}) \dots W^1(\bm{\theta}_1) E(\bm{x}) W^0(\bm{\theta}_0)$ is the quantum circuit with repeated $L$ layers, and $O$ is an observable. \citet{Schuld_2021} showed that these type of models can be represented as: 
\begin{equation}\label{eq:fourier_series}
    f_{\bm{\Theta}}(\bm{x}) = \sum_{\bm{\omega} \in \Omega} c_{\bm{\omega}} e^{-i \bm{\omega} \bm{x}},
\end{equation}
where $\Omega$ is the frequency spectrum and $c_{\omega}$ are the coefficients. One of the interesting findings of the study~\cite{Schuld_2021} is that as the number of times data is re-uploaded into the model increases (reflected by a higher number of layers $L$), the more the set of frequencies $\Omega$ available to the model grows. This in turn enables the quantum mode $f_{\bm{\Theta}}(\bm{x})$ to express increasingly more complex functions.

\subsection{Fourier-based classical surrogates}
\label{sec:surrogates}

From \cref{eq:fourier_series} it follows that we can derive a fully classical representation $s_{\bm{c}}(\bm{x}) = \sum_{\bm{\omega} \in \Omega} c_{\bm{\omega}} e^{-i \bm{\omega} \bm{x}} \approx f_{\bm{\Theta}}(\bm{x})$, where $\bm{c} = (c_{\omega})_{\omega \in \Omega}$ are the coefficients that need to be optimized to closely replicate the output of target model $f_{\bm{\Theta}}(\bm{x})$. More rigorously, \citet{Schreiber_2023} define the \textit{classical surrogates} $s_{\bm{c}}$ using Probably Approximately Correct (PAC) framework as follows:

\begin{definition}[Classical surrogate]
A classical surrogate $s$ belongs to a hypothesis class of quantum learning models $\mathcal{F}$ if there exists a conversion (surrogation) process that transforms $f \in \mathcal{F}$ into $s$ such that:
\begin{equation}
    \mathbb{P}\big[ sup_{\bm{x} \in \mathcal{X}} ||f_{\bm{\Theta}}(\bm{x}) - s_{\bm{c}}(\bm{x})|| \leq \epsilon \big] \geq 1 - \delta,
\end{equation}
where $\epsilon$ is the error bound and $\delta$ is the failure probability. It is required that the \textit{surrogation process} is efficient in the quantum model size.
\end{definition}

In other words, $s_{\bm{c}}(\bm{x})$ is called a classical surrogate if it matches the predictions of $f_{\bm{\Theta}}(\bm{x})$ closely enough with high enough probability on the dataset $\mathcal{X}$. The supremum norm ensures that even the outliers of $s_{\bm{c}}(\bm{x})$ do not deviate too far from $f_{\bm{\Theta}}(\bm{x})$. Additionally, the process of creating $s_{\bm{c}}(\bm{x})$ is required to be efficient, a point that we critically examine in this work.

\subsection{Surrogation process}\label{sec:surrogation}
\citet{Schreiber_2023} proposed the following process from creating classical surrogates $s_{\bm{c}}$ from quantum models $f_{\bm{\Theta}}$:

\subsubsection{Grid generation} For each feature $i$, we sample a set $T_i$ of equidistant points in the interval $[0, 2 \pi )$ and generate a grid $T$ that consists of all possible combinations of these points. The number of points sampled in this interval depends on maximal frequency $\omega_{max}(i)$ of the given feature $i$ and is calculated as $T_i = 2 \omega_{max}(i) + 1$. The total grid size is then determined as $T = \prod^d_{i=1} T_i$, which is governed by the width of the quantum model that influences $d$, the depth of the model that controls $\omega_{max}(i)$~\cite{Schuld_2021}, as well as the size of the interval, in which the points are sampled. 

\subsubsection{Circuit sampling} For each point in the grid $x_j \in T, j \in [1, |T|]$, we acquire a quantum model output $\bm{\hat{y}} = f_{\bm{\theta}}(\bm{x})$, which corresponds to the expectation values of the quantum circuit. The computational costs of this step depends on the size of the grid $|T|$ and the number of circuit calls from which  the expectation values are calculated.

\subsubsection{Solving the system of linear equations} \label{sec:surrogation}
To find an optimal setting of the Fourier coefficient $\bm{c}$ (see \cref{sec:surrogates}), we can solve a linear system: 
\begin{equation}\label{eq:lin_system}
    \bm{c}^* = \arg\!\min_{\bm{c}} || A \bm{c} - \bm{\hat{y}} ||^2,
\end{equation}
where
\begin{equation}\label{eq:A}
A =
\begin{bmatrix}
e^{-i \omega_1 x_{1}} & e^{-i \omega_2 x_{1}} &  \cdots & e^{-i \omega_{max} x_{1}} \\
e^{-i \omega_1 x_{2}} & e^{-i \omega_2 x_{2}} & \cdots & e^{-i \omega_{max} x_{2}} \\
\vdots & \vdots & \ddots & \vdots \\
e^{-i \omega_1 x_{|T|}} & e^{-i \omega_2 x_{|T|}} & \cdots & e^{-i \omega_{max} x_{|T|}}
\end{bmatrix}
\end{equation}
This surrogation process scales sublinearly in number of quantum circuit executions~\cite{Schreiber_2023}.

\citet{landman2022classicallyapproximatingvariationalquantum} introduced a classical representation for variational quantum circuits using the Random Fourier Features (RFF) method, which was initially developed for approximating large kernels. This approach differs from the method proposed in~\cite{Schreiber_2023}, as it involves randomly sampling a subset of frequencies rather than calculating the entire set. This methods is not exact like~\cite{Schreiber_2023}, but it delivers probabilistic guarantees of recovery. The author showed the inherent redundancies of the frequencies spectrum that allow the utilization of only a fraction of them. They propose three RFF sampling strategies: \textit{distinct sampling}, \textit{tree sampling} and \textit{grid sampling}. The validity of this method has been shown in a simulation environment on a small scale, utilizing up to 5 qubits. In contrast, our work demonstrates that by increasing the scale of quantum models it exposes the inefficiencies of the RFF method if used as a standalone solution in practice. We test proposed methods that are not embedding strategy specific, such as \textit{distinct sampling}, and introduce our adaptation.

\section{Surrogation process 2.0}\label{sec:surrogation_2.0}
There is a significant caveat of the procedure described above that prohibits its application for quantum models $f_{\bm{\Theta}}(\bm{x})$ of any reasonable size. The memory requirement for it, which involves storing the matrix from \cref{eq:A}, increases as $(2 \omega_{\max} + 1)^{|T|}$. Practically, this means that the available resources of classical devices limit the complexity of $f_{\bm{\Theta}}(\bm{x})$ we can convert into $s_{\bm{c}}$, as illustrated in \cref{tab:ram_requirement}. For example, with just a 2-layer model, we can at maximum represent a 13-qubit model and we will require access to a High Performance Computing (HPC) system. This falls significantly short of industry-relevant scales. In the following, we highlight the redundancies in the procedure that lead to excessive computational memory requirements and propose an alternative method that substantially reduces the computational resources required. The two key adjustments are listed below.
\begin{table}[h]
    \centering
    \begin{tabular}{|l | r | c | c | c |}
        \hline
          \multicolumn{2}{|c|}{Device}  & \multicolumn{3}{c|}{Maximal number of qubits}\\
          \hline
          \hline
         Level & RAM & 1 layer & 2 layers & 3 layers \\ 
         \hline
         Laptop & 16 GB & 6 - 7 & 4 & 2 - 3 \\
         Workstation & 8 TB & 13 & 7 & 5 \\
         HPC & 1.5 PB & 26 & 13 & 8 - 9 \\
         \hline
    \end{tabular}
    \caption{Necessary RAM required to store a large matrix $A$ from \cref{eq:A}, the class of classical devices needed, and the approximate number of qubits and layers in a quantum model for which a classical surrogate can still be generated}
    \label{tab:ram_requirement}
\end{table}

\subsection{Dataset instead of a full grid}\label{sec:dataset_vs_grid}

\begin{figure}
    \centering
    \includegraphics[trim=2.5cm 2cm 2.5cm 2.75cm, clip, width=0.5\textwidth]{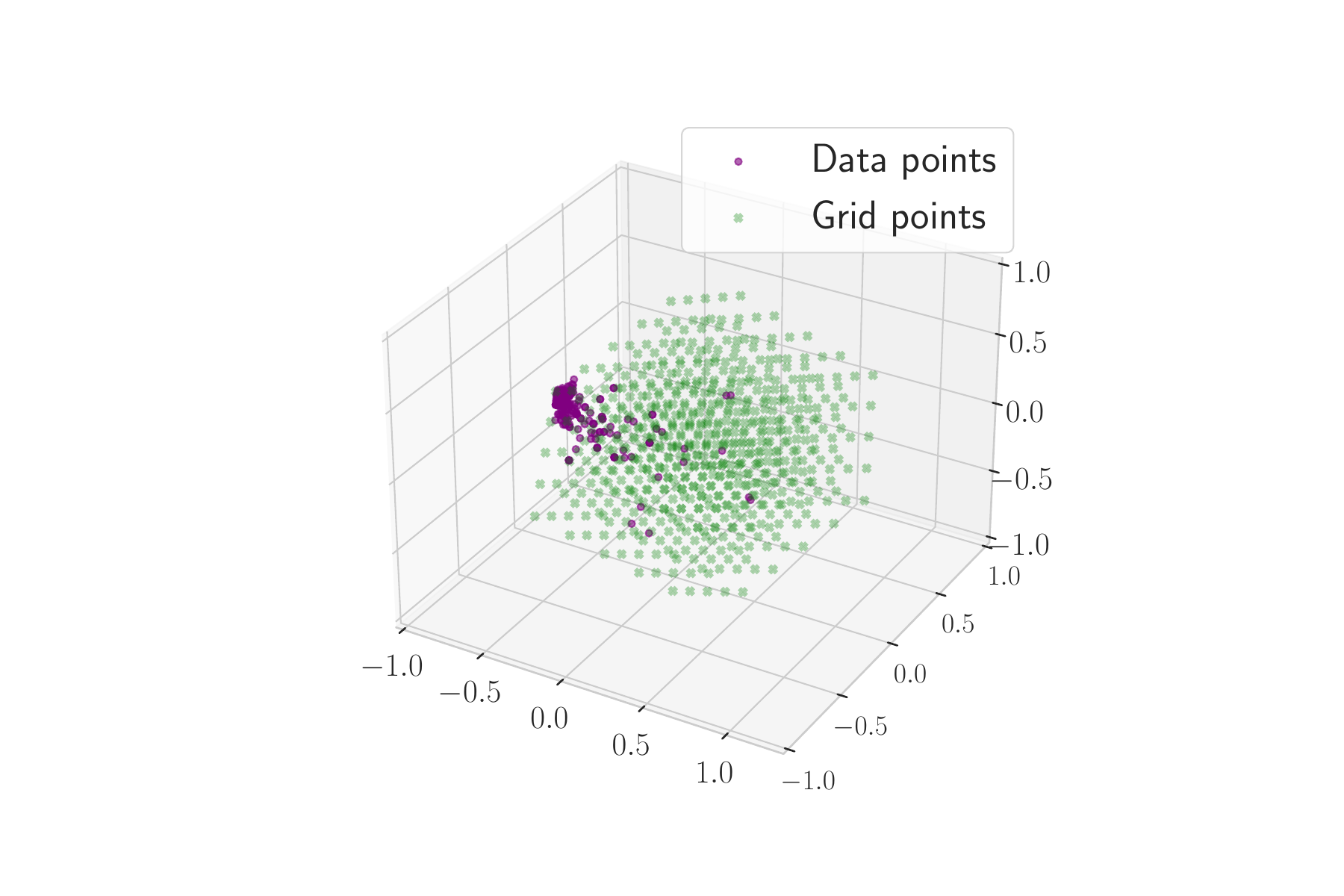}
    \caption{Visualization of the volume occupied by the actual training data within an extensive sampling grid from \cref{sec:surrogation}.}
    \label{fig:grid_vs_training}
\end{figure}

An initial grid range in \cref{sec:surrogation} was proposed to extract the complete Fourier spectrum in a dataset-agnostic manner. While this approach offers guarantees of the identity between quantum and classical outputs~\cite{Schreiber_2023}, it also encompasses a significant amount of practically irrelevant information, representing the overwhelming majority of the extracted data. From an application viewpoint, achieving extremely high precision in duplication of quantum model's behaviour on any possible dataset is not essential\footnote{Assuming that the quantum model has been trained already and that a quantum advantage is expected during training.}. However, it is crucial to significantly reduce memory requirements to go beyond quantum model sizes listed in \cref{tab:ram_requirement}. Given that the output of the quantum model has been optimized on the available dataset alone during the training process, it is safe to assume that we can replace the entire grid with just the training data (guarantees of this method are discussed in \cref{sec:guarantess}). \cref{fig:grid_vs_training} illustrates the amount of redundancy included when employing the full grid compared to using just the dataset. The particular dataset description used for our experiments is included later in \cref{sec:Dataset}.

\subsection{Random frequencies sampling}\label{sec:frequency_sampling}

\begin{figure}
    \centering
    \includegraphics[width=0.45\textwidth]{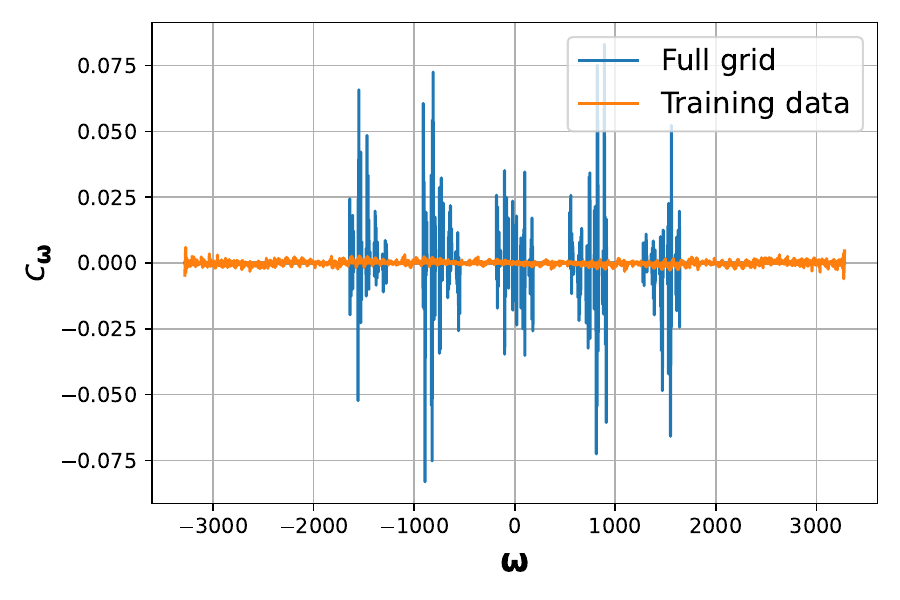}
    \caption{Illustration that the redundancy of frequency increases when only the data points are considered in place of a full grid}
    \label{fig:frequencies_grid_vs_dataset}
\end{figure}

Similar to grid considerations above, the initial surrogate proposal assumes utilization of the entire Fourier frequency spectrum. However, as demonstrated in the work by \citet{landman2022classicallyapproximatingvariationalquantum}, significant redundancies exist within this spectrum as well, leading to unnecessary computational overhead. In our experiments, we rely on one of the proposed sampling strategies in~\cite{landman2022classicallyapproximatingvariationalquantum}, specifically distinct sampling. This approach allows us to randomly sample a small subset of frequencies. Interestingly, we demonstrate that the fraction of redundant frequencies increases even further when we substitute the grid with the dataset, as illustrated in \cref{fig:frequencies_grid_vs_dataset}. This allows us to reduce memory requirement even further.

\subsection{Summarizing the algorithm}
\cref{alg:surrogation} provides an overview of the entire surrogation pipeline from \cref{sec:surrogation}, which incorporates the modifications proposed in the preceding sections. It highlights the key adjustments that enhance the performance and applicability of the surrogates.

\begin{algorithm}[h]
\caption{Surrogation process 2.0}\label{alg:surrogation}
\begin{algorithmic}
\Require $f_{\bm{\theta}}$ is trained
\Ensure $f_{\bm{\theta}} \approx s_{\bm{c}}(\bm{x}) = \sum_{\bm{\omega} \in \Omega} c_{\bm{\omega}} e^{-i \bm{\omega} \bm{x}}$

\State $T \gets \mathcal{X}$ \Comment{\cref{sec:dataset_vs_grid}}
\State $\Omega = \{\bm{w}_1, \dots, \bm{w}_{|\Omega|}\} \gets U(-L, L)$ \Comment{\cref{sec:frequency_sampling}}

\For{$\bm{x}_i \in T$}
\State $\bm{\hat{y}} \gets f_{\bm{\theta}}(\bm{x}_j)$
\For{$\bm{w}_j \in \Omega$}
\State $A_{ij} \gets e^{-i \bm{w}_j \bm{x}_i}$
\EndFor
\EndFor
\State $\bm{c}^* \gets \arg\!\min_{\bm{c}} || A \bm{c} - \bm{\hat{y}} ||^2$

\end{algorithmic}
\end{algorithm}

\section{Experimental setup}\label{sec:method}
To validate the practicality of the proposed surrogation process outlined in \cref{alg:surrogation}, we conducted an empirical study that is described below.

\subsection{Use-case}\label{sec:Dataset}

For our empirical studies, we selected a dataset collected from one of E.ON's Combined Heat and Power plants. These type of plants generate both electricity and thermal energy and they are commonly used in industrial processes, where multiple energy source can be utilized. It consists of historical recordings from $53$ sensors over $2179$ time steps. The task is to predict the required energy output of the power plant (in $kW$) to meet customer's demand. The dataset has been normalized and rescaled to $[0,1]$. The Density-Based Spatial Clustering of Applications with Noise (DBSCAN) method~\cite{10.5555/3001460.3001507} is applied to eliminate outliers and noise by identifying high-density regions as healthy clusters while classifying data instances that fall outside these regions as anomalies. Additionally, the dataset dimensions have been reduced for various input conditions using Principal Component Analysis (PCA)~ \cite{10.1145/3447755}.

\subsection{Model}

\begin{figure}[h]
    \centering
    \includegraphics[width=0.5\textwidth]{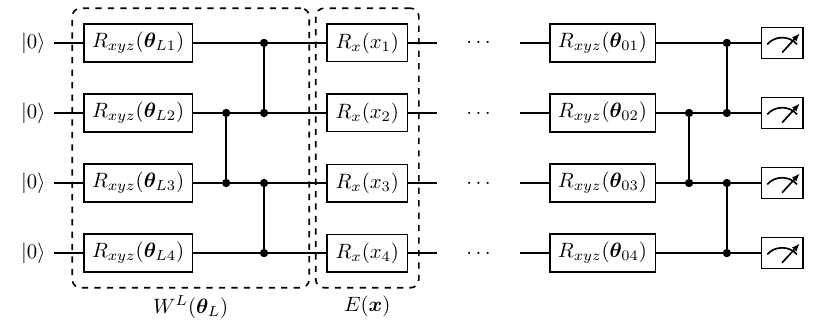}
    \caption{A 4-qubit illustration of a variational quantum circuit that follows the same schematics as \cref{fig:reuploading}. This quantum model is trained to perform regression on the dataset (\cref{sec:Dataset}) and its classical surrogate is created.}
    \label{fig:quantum_architecture}
\end{figure}

The quantum learning model $f_{\bm{\theta}}$ is represented by a quantum reuploading architecture as discussed before (see \cref{fig:reuploading})\footnote{This approach is applicable to a broader QNN architecture as it can also be considered as a special case of a reuploading architecture - with a single layer}. The general architecture was inspired by the model from the original paper of classical surrogates~\cite{Schreiber_2023} and consists of the following elements. The parametrized layers $W(\bm{\theta})$ consists of $R_{xyz}(\bm{\theta})$ gates followed by entanglement layers designed to align with the coupling map of the chosen quantum chip - \texttt{ibm\_kyiv}.\footnote{The SWAP operation bridges connectivity gaps between unconnected qubits, but it is an expensive operation (it requires three successive CNOT gates). The entangling strategy chosen for this architecture proved to significantly reduce the depth of the circuits in our internal experiments.} The embedding layer $E(\bm{x})$ consists of an Angle embedding with $R_{x}(\bm{x})$ gates, which is a popular choice within a data reuploading framework \cite{ Schuld_2021, Schreiber_2023}. The output of a circuit is postprocessed to attain expectation values of each qubit, which are then averaged over to gain a single output value. A sketch of the entire architecture is illustrated in \cref{fig:quantum_architecture}. 

\subsection{Implementation}

\subsubsection{Model training}
The quantum circuit is implemented in the \texttt{Qiskit (v 1.3.1)} framework~\cite{qiskit2024} and is trained using the built-in functionality of \texttt{EstimatorQNN} primitive from the \texttt{qiskit\_machine\_learning} framework. We chose the \texttt{COBYLA} optimizer for our experiments as it showed fast convergence rates in previous works. For the hardware runs, we selected \texttt{ibm\_kyiv}, which offers 127 qubits. For noisy simulated runs, we use \texttt{qiskit-aer (v 0.16.0)} simulator, in which we load a noise profile of \texttt{ibm\_kyiv} machine.
\subsubsection{Surrogation process}
To solve \cref{eq:lin_system}, we employ the Moore-Penrose pseudoinverse method that utilizes a Singular Value Decomposition implemented in \texttt{scipy.linalg.pinv} package. This is a computationally light and stable method for solving systems of linear equations.

\section{Results}\label{sec:results}
The surrogation process has two key requirements: resource efficiency and performance reliability.
Below, we detail the empirical results of the proposed method, assessing its performance across various conditions, including different scales and different simulation environments.

\subsection{Performance}

\begin{figure}
    \centering
    \includegraphics[width=0.45\textwidth]{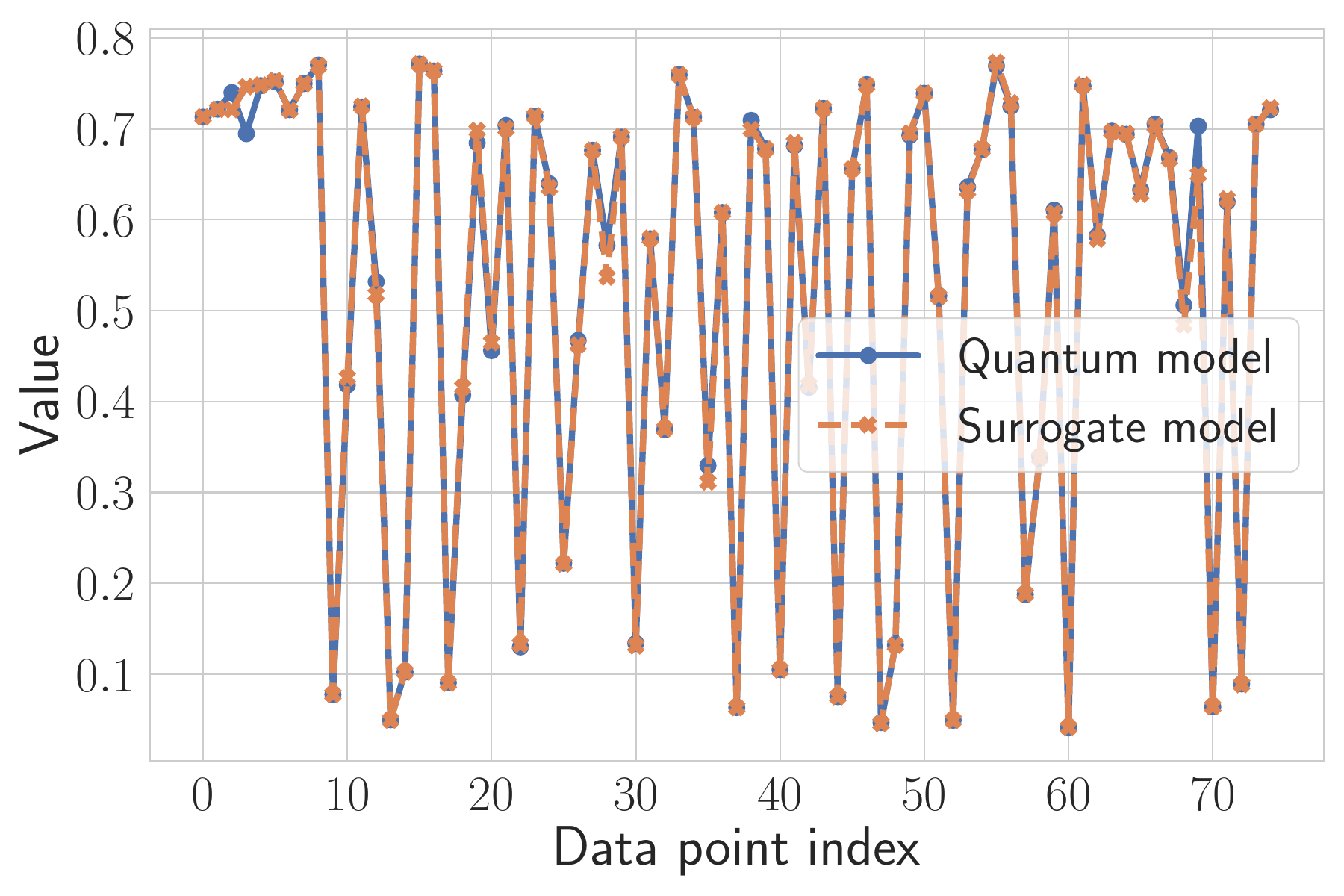}
    \caption{Predictive performance comparison between a 9-qubit 2-layer quantum model and its classical surrogate on test dataset, while utilizing only $0.3 \times 10^{-9}\%$ of frequencies.  For clarity, only a subset of the test set performance is presented.}
    \label{fig:9_qubit_showcase}
\end{figure}

We present the performance of a proof-of-concept implementation. The ability to create a classical surrogate for a model of this size represents a significant advancement beyond previous possibilities shown in \cref{tab:ram_requirement}. \cref{fig:9_qubit_showcase} demonstrates a 9-qubit 2-layer quantum learning model $f_{\bm{\Theta}}(\bm{x})$ alongside its classical surrogate $s_{\bm{c}}(\bm{x})$, created by utilizing the method described in \cref{alg:surrogation}. The training of $f_{\bm{\Theta}}(\bm{x})$ was performed in a noiseless simulation environment, it was then converted into a $s_{\bm{c}}(\bm{x})$ on a laptop with 16 GB RAM. Both models were later tested on the testing dataset. Our results demonstrate that, with just a small fraction of $0.3 \times 10^{-9}\%$ of the available frequencies, it is possible for $s_{\bm{c}}(\bm{x})$ to achieve a Mean Squared Error (MSE) of $0.0224$. This empirically confirms the previous assertion about redundancies being included into the prior proposed process. 

\subsection{Minimizing computational demand}

\begin{figure}[t]
    \centering
    \begin{subfigure}[t]{0.45\textwidth}
        \centering
    \includegraphics[width=\textwidth]{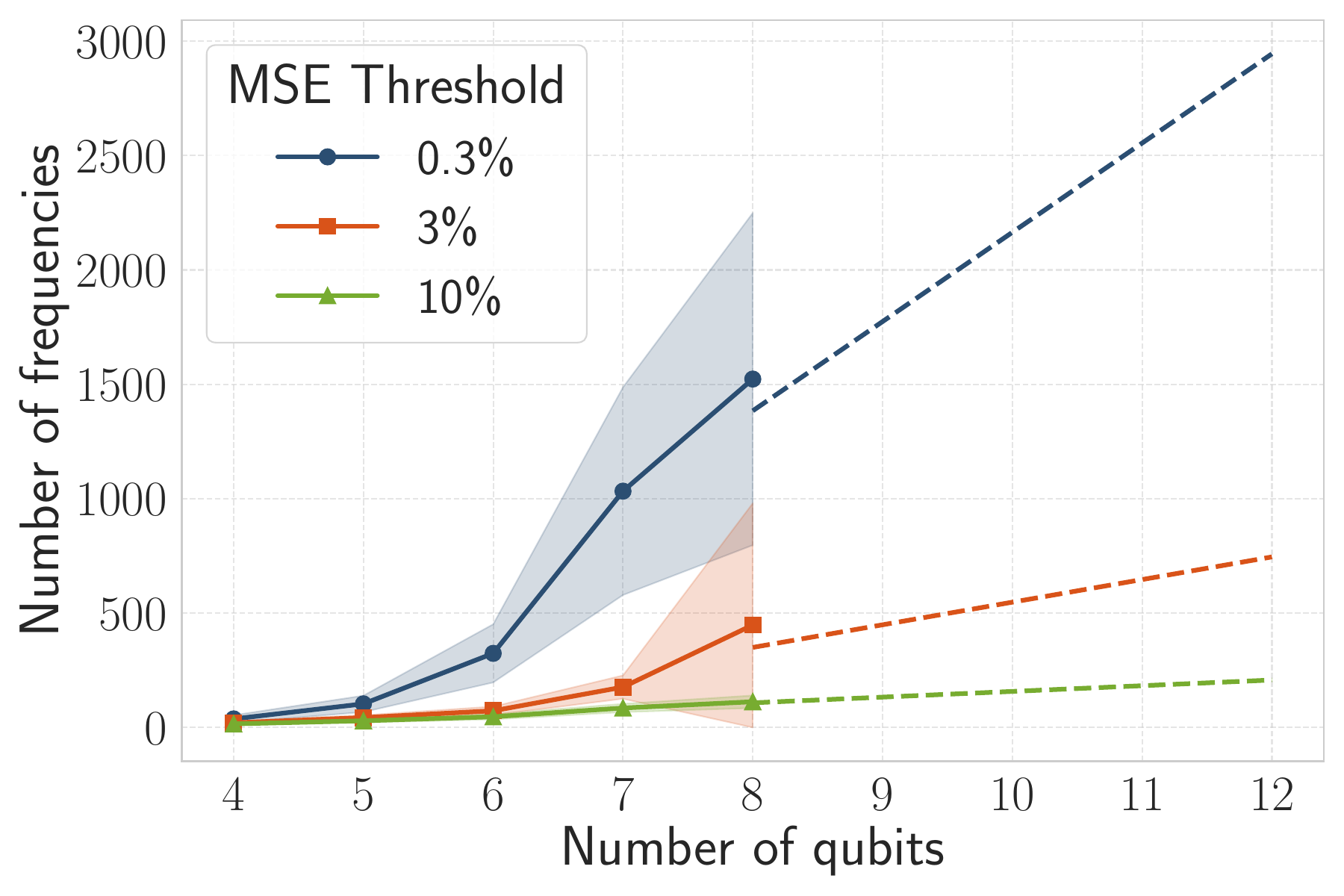}
        \caption{Frequency study. The number of data points was set to the size of the entire dataset.}
        \label{fig:frequencies_requirement}
    \end{subfigure}%
    
    \begin{subfigure}[t]{0.45\textwidth}
        \centering
    \includegraphics[width=\textwidth]{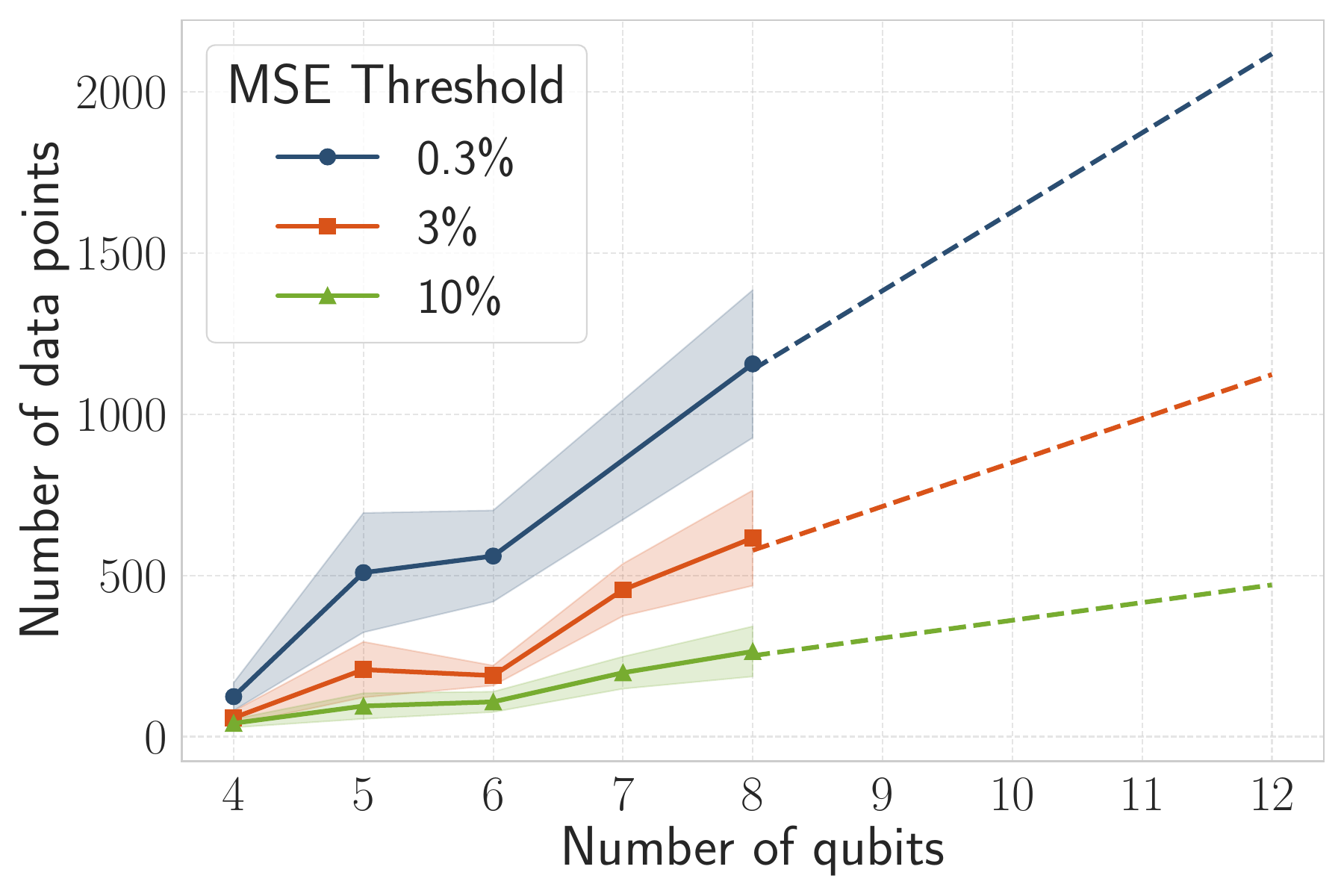}
        \caption{Data points study. The number of frequencies is set to 10k.}
        \label{fig:training_requirement}
    \end{subfigure}
    \caption{Required number of a frequencies and datapoints to achieve desired MSE precision between quantum model and its classical surrogate that depends on number of qubits.}
\end{figure}

In industrial settings, a perfect classical replica of a quantum model is not always necessary, allowing us to relax the demand for a low MSE score. This flexibility allows to further reduce computational resource demand. To explore this scenario, we examined three different MSE threshold values deviations (0.3\%, 3\% and 10\%), analyzing the corresponding number of frequencies (\cref{fig:training_requirement}) and data points (\cref{fig:training_requirement}) necessary to meet each threshold. 
The data presented in the plots was obtained by executing \cref{alg:surrogation} for quantum models from 4 to 8 qubits\footnote{The choice to restrict the plots to 8 qubits is based on the dataset size, a challenge discussed below. Conducting experiments with more than 8 qubits requires augmenting the dataset. An example of a larger-scale experiment can be found in \cref{sec:appendix_dataset}.} wide for over 20 iterations to ensure statistical significance. By varying the numbers of frequencies or datapoints and tracking achieved MSE by $s_{\bm{c}}$, we established the minimal average number of these quantities necessary to consistently meet the specified threshold. We then fitted \texttt{LinearRegressor} to predict requirements for model sizes beyond what was tested. 

As expected, the plots reveal higher demand and higher deviation on both quantities under stricter performance requirements (lower MSE deviation). However, the required resources appear to scale only linearly, whereas in the original proposal they scaled exponentially. This behavior aligns with theoretical predictions; in \cref{sec:guarantess}, we demonstrate that the theoretical guarantees also scale linearly, thereby confirming our empirical observations. This linear scaling arises from our implementation's reliance on the random Fourier features method that was adapted from \cite{landman2022classicallyapproximatingvariationalquantum}. These results indicate a significant improvement in the feasibility of applying the surrogation method at industry-relevant scales.

However, a minor bottleneck becomes apparent: as the demand for the number of data points increases linearly with the size of the model, it may surpass the available data points. In \cref{sec:appendix_dataset}, we discuss strategies for augmenting the existing dataset, which a well-explored area within the ML field. Additionally, we observe that the variance in frequency demand increases significantly as the system grows, indicating the presence of advantageous and disadvantageous subsets of frequencies. 

\subsection{Accounting for hardware noise}

Running our method on a noisy device will require additional resources, as the output of a noisy quantum circuit cannot be easily represented by a Fourier series. To better assess the noise impact, we conduct several experiments: scaling experiments with noisy \texttt{Qiskit} simulators, validation of the results on \texttt{ibm\_kyiv} quantum hardware, and testing the effectiveness of out-of-the-box error mitigation techniques.

\subsubsection{Noisy simulators} First experimental results were acquired from noisy simulators. \cref{fig:noisy} represents how resource demand explodes in the presence of noise. Due to this increased computational demand of the simulation the experiment could only be conducted for a limited number of qubits. The observed trend suggests that the complexity introduced by noise may scale exponentially with the number of qubits, highlighting the bottleneck of the surrogation of noisy quantum circuits and the need to explore error mitigation.

\subsubsection{Quantum hardware} 
\begin{figure}
    \centering
    \includegraphics[width=0.4\textwidth]{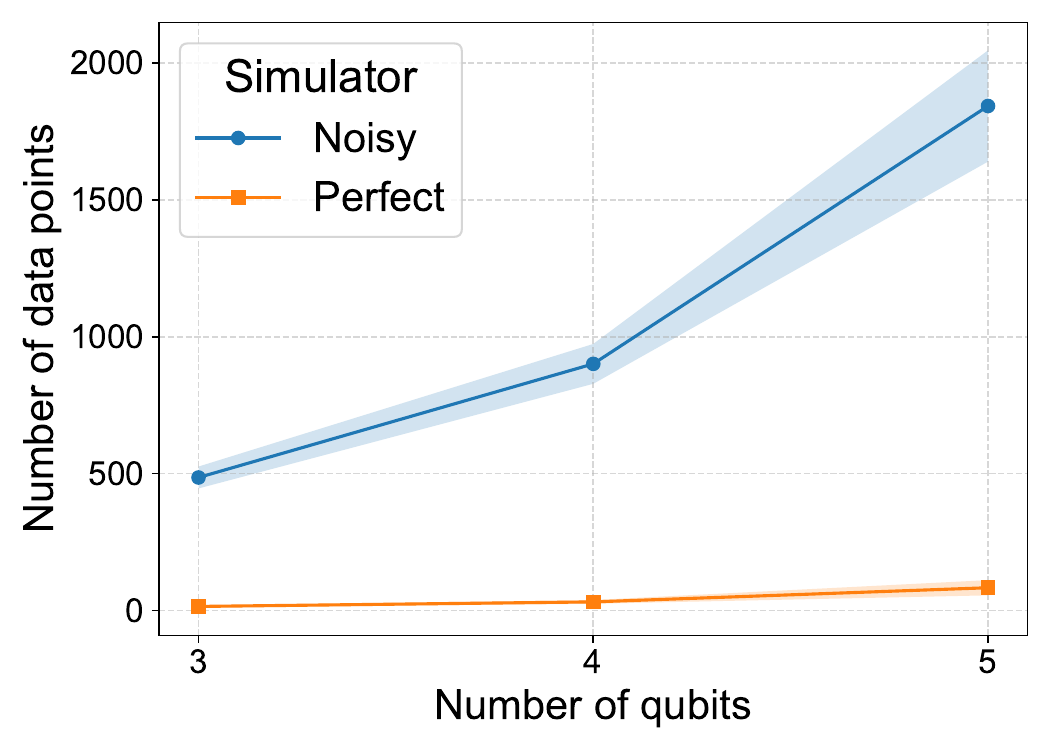}
    \caption{Required number of datapoints to achieve at most $10\%$ MSE deviation between quantum model and its classical surrogate that depends on number of qubits and presence of noise.}
    \label{fig:noisy}
\end{figure}

We then proceeded to the hardware experiments. Given that the trainability of quantum circuits on noisy hardware was not the primary focus of this study, and considering the high cost associated with training on quantum hardware, we performed a small case study. To achieve this, we pretrained a 4-qubit circuit using a noisy simulator before transferring the training process to the quantum hardware for the remaining steps. After a warm-start on the simulators, we trained the quantum model on 28 points from the training dataset over 10 iterations with the COBYLA optimizer, achieving MSE of $0.009$ on 42 points of testing dataset. We then proceeded to creating a classical surrogate from this model. By utilizing 1,131 training data points to sample the trained quantum model across all frequencies, we achieved a MSE of $0.013$, which is only marginally higher than observed performance from the quantum models. This result demonstrates that our approach remains effective even on noisy quantum hardware, and that selecting the number of data points in accordance with \cref{fig:noisy} helped maintain a low error rate for the surrogate. However, these findings also confirm the increased computational demands imposed by noise, highlighting the need for further investigation into error mitigation techniques.

\subsubsection{Error mitigation} As the next step we explored the effect of error mitigation techniques on data demand of our approach. 
We repeated the experiment from above, but this time employing \texttt{resilience level} 2, integrating \texttt{Zero Noise Extrapolation (ZNE)} and \texttt{gate twirling}. In this case the model was warm-started on noiseless simulator. While keeping the same number of resources as in the previous experiment, the quantum model achieved an MSE of $0.0177$, while surrogate slightly outperformed the model achieving MSE of $0.0164$. This suggest that even out-of-the-box approaches for error mitigation are capable to effectively compensate for noise-induced accuracy degradation, potentially reducing the data requirement of the algorithm. The caveat is of course the cost of quantum resources required to execute these algorithms.

\section{Discussion}\label{sec:discussion}
In our study, we exposed a prohibiting computation demand of the surrogation process and proposed a method to avoid computational redundancies, bringing quantum solutions closer to practical use in industrial environments. The empirical validation of our approach showed strong results. However, in our experiments, we focused on a single model architecture, which may not capture the full range of possibilities available in the field. Additionally, our analysis relied on a single dataset, which restricts the generalizability of our findings. While we expect the performance of our approach to vary with different models and datasets, we believe the trends identified will remain consistent. Future research should focus on validating our approach across a wider range of conditions and at larger scales. We faced a bottleneck that limited our ability to conduct a more extensive empirical study on quantum devices, particularly due to trainability issues on noisy devices associated with models spanning large Hilbert spaces. Insights from the literature, such as \cite{Thanasilp_2023} and \cite{Wang_2021}, may help address these challenges in future work.

Our empirical results revealed a lot of deviations in performance during the frequency study, indicating that certain subsets of frequencies are more critical than the others. The importance of frequencies potentially depend on the architectural qualities of trained model, and studies like \cite{10821430, Sakhnenko_2022} can provide further insights. Understanding this aspect in future work could substantially reduce the computational complexity of the method even further, paving the way for larger classical surrogates.

The existence of classical representations of quantum models imposes significant limitations on the potential sources of quantum advantage. \citet{Schreiber_2023} suggests that any quantum advantage for models that can be represented by classical surrogates can manifest solely during the training phase. With the method proposed in our work, it becomes feasible to perform empirical study at a meaningful scale to investigate when practical advantages can be realized through e.g. enhanced trainability. Numerous studies have already explored trainability of quantum models, which provides a solid foundation for further research~\cite{Thanasilp_2023, Barthe2024gradientsfrequency, gilfuster2025relationtrainabilitydequantizationvariational}. However, a more solid theoretical understanding and empirical validation of this potential advantage is still lacking, making it an essential direction of future research.

Another type of research provides alternative ways of \textit{dequantifying} (representing classically) quantum models, such as shadow models~\cite{Jerbi_2024} and tensor networks~\cite{berezutskii2025tensornetworksquantumcomputing}. For future studies it is important to consider the broad spectrum of these techniques and examine their interconnections. A potential quantum advantage may arise from the limitations of these methods, in scenarios where models can no longer be efficiently dequantified.

\section{Conclusion}\label{sec:conclusion}
The ability to represent QML algorithms classically is particularly intriguing from an application standpoint, as it eliminates the necessity for on-demand access to quantum hardware. This is especially relevant for industries with specific use cases, including: (1) Real-time applications, where cloud access can introduce latency, such as in edge and IoT devices; (2) Safety-critical sectors, where cloud access may pose security risks, such as in energy, healthcare, and defense; and (3) High volumes of requests, where dependence on cloud access can become prohibitively expensive. In this work, we identify and address the prohibitively high computational demands of generating these classical representation, known as classical surrogates~\cite{Schreiber_2023}, by proposing an adapted surrogation routine. This research paves the way for the accelerated integration of QML approaches in industrial settings and enhances the pursuit of practical quantum advantages in empirical applications. We demonstrate the effectiveness of our method on a real-world energy demand forecasting problem, conducting rigorous testing of performance and computational demand in both simulations and on quantum IBM hardware. We demonstrate a proof-of-concept implementation of our approach that was able to transform quantum models that would have required TBs of RAM on just a standard laptop, significantly downscaling the algorithmic space complexity. Furthermore, our results indicate that our method achieves high accuracy on the testing dataset while its resource requirements scale linearly rather than exponentially. Our work represents a significant step towards utilizing QML algorithms in real-world applications.

\printbibliography

\begin{appendices}
\section{Guarantees}\label{sec:guarantess}
The original surrogation method \cite{Schreiber_2023} utilized an entire grid of all possible input combinations, resulting in significant memory complexity. This approach guaranteed that the reconstruction error would remain within specific error bounds. In contrast, our method presents a more practical alternative, however, it is an approximate method, which undermines the applicability of the original error bounds. Here, we provide theoretical bounds for the number of frequency samples necessary to guarantee a certain error between the quantum model and the surrogate applicable to our method.

The goal is to bound the error $\epsilon$ between the quantum model $k$ and its approximation (classical surrogate) $\tilde{k}$ in a training method agnostic way for a given number of frequency samples $D$:

\[||f||_\infty=
||k(x)-\tilde{k}(x)||_\infty \le \epsilon
\]
with probability of at least $1-\delta$ over the domain $X$, similarly to \citep{sutherland2015errorrandomfourierfeatures}. We can analyze surrogate approximation properties through the prism of kernel theory by represent the PQC using a continuous shift-invariant kernel function \[ k : X \times X \to \mathbb{R}. \] For the specific embedding we can even define the PQC as follows:
\begin{equation}
  {k}(x, x') = \frac{1}{D}\sum_{\omega \in \Omega} \cos\bigl(\omega\cdot (x - x')\bigr)
\end{equation}
From \citep{landman2022classicallyapproximatingvariationalquantum}, the surrogation process approximates this kernel by averaging over $D$ randomly drawn frequencies:
\begin{equation}
\tilde{k}(x, x') = \tilde{\phi}(x)^T \tilde{\phi}(x').
\end{equation}


Following the argumentation line from \cite{sutherland2015errorrandomfourierfeatures}. Assuming $X$ is compact with diameter $\ell$, we denote $k$'s Fourier transform as $P(\omega)$, $\sigma_p^2 = \mathbb{E}_p \left[ \|\omega\|^2 \right]$. 
For any $\varepsilon > 0$, let

\begin{equation}
\label{alpha_epsilon}
\alpha_\varepsilon := \min\left(1, \sup_{x,y \in X} \frac{1}{2} + \frac{1}{2}k(2x, 2y) - k(x, y) + \frac{1}{3}\varepsilon \right)
\end{equation}

\begin{equation}
\label{beta_d}
\beta_d := \left( (\frac{d}{2})^\frac{-d}{d+2} + (\frac{d}{2})^\frac{2}{d+2}\right) 2^\frac{6d+2}{d+2}
\end{equation}

\par
we then assume that
\begin{equation}
\label{dataset_requirement}
\epsilon \le \sigma_pl
\end{equation}
which leads to the following error bound:
\begin{align}
\label{max_error_probability}
Pr\Biggl(\sup_{x,y\in X}\bigl|& k(x-y) - \tilde{\phi}(x)^T\tilde{\phi}(y) \bigr| \ge \epsilon\Biggr) \le \nonumber\\ & \beta_d \left(\frac{\sigma_pl}{\epsilon}\right)^\frac{2}{1+\frac{2}{d}} \exp\Biggl(-\frac{D\epsilon^2}{8(d+2)\alpha_\epsilon}\Biggr),
\end{align}
From which devite the following \citep{sutherland2015errorrandomfourierfeatures}:
\begin{equation}
\label{max_error_bound}
D \geq \frac{8(d + 2)\alpha_\varepsilon}{\varepsilon^2} \left(\frac{2}{1+\frac{2}{d}} \log \frac{ \sigma_p \ell}{\varepsilon} + \log \frac{\beta_d}{\delta} \right). 
\end{equation}
This bound grows linearly in dimension $d$, meaning that even for large circuits with tens of qubits, only a linear increase in the required frequency samples can be expected.

We can further tighten the bound in Equation \ref{max_error_bound}, following the reasoning from \citep{landman2022classicallyapproximatingvariationalquantum}, by focusing on the specific case studied in this work: the Pauli encoding scheme and linear ridge regression as the training method. In the case of Pauli encodings, each gate contributes eigenvalues of  \(\pm \frac{1}{2}\), yielding a frequency spectrum \(\Omega\) made up from these eigenvalues. Consequently, the PQC model \(k(x)\) is expressed as
\[
k(x) = \sum_{\omega} \bigl(a_{\omega} \cos(\omega x) + b_{\omega} \sin(\omega x)\bigr).
\]
This kernel can approximate by randomly select \(D\) samples from \(\Omega\) forming a random Fourier feature as follows:
\[
\tilde{\phi}(x) = \frac{1}{\sqrt{D}}
\begin{bmatrix}
  \cos(\omega_1^T x) \\[1mm]
  \sin(\omega_1^T x) \\[1mm]
  \vdots \\[1mm]
  \cos(\omega_D^T x) \\[1mm]
  \sin(\omega_D^T x)
\end{bmatrix}.
\]

In case of a Linear Ridge Regression (LRR), we consider a training set \(\{(x_i,y_i)\}_{i=1}^M\) and define:
\begin{itemize}
  \item \( f \): the LRR model trained using the true kernel \( k \) with regularization \( \lambda = M\lambda_0 \) (for some \( \lambda_0 > 0 \));
  \item \( \tilde{f} \): the LRR model trained with the approximate kernel \( \tilde{k}(x,x')=\tilde{\phi}(x)^T\tilde{\phi}(x') \) under the same regularization.
\end{itemize}
Then taking into consideration the Pauli embedding, the prediction error can be bounded with high enough probability (at least \( 1-\delta \)) as
\[
|f(x)-\tilde{f}(x)| \le \epsilon,
\]
provided that the number of random features \( D \) satisfies
\begin{align}
\label{final_bound}
D = \Omega\!\Biggl( d\, C_1\frac{(1+\lambda)^2}{\lambda^4\epsilon^2}\Bigl(&\log(dL^2|X|)+\nonumber\\&\log\!\Bigl(C_2\frac{(1+\lambda)}{\lambda^2}-\log\delta\Bigr)\Bigr) \Biggr),
\end{align}
where \( C_1 \) and \( C_2 \) are constants that depend on $\sigma_y^2 = \frac{1}{M}\sum_{i=1}^{M} y_i^2$ and \( |X| \) \cite{landman2022classicallyapproximatingvariationalquantum}.

This bound was derived in a slightly different context by \cite{landman2022classicallyapproximatingvariationalquantum}. Specifically, we utilize ordinary least squares to derive Fourier coefficients instead of linear ridge regression (LRR), which includes regularization of these coefficients—a step we have omitted. As a result, our surrogate model may capture finer-grained behavior from the quantum outputs but is also more sensitive to noise \citep{englesson2021consistencyregularizationimproverobustness}. Nevertheless, one could apply our method using LRR and reasonably expect the tighter bound to hold, although we have not extensively tested this approach due to the slow convergence of LRR in certain scenarios.

Generally, this bound confirms that, while the frequency spectrum of the PQC can be exponentially large, only a relatively small subset significantly contributes to the model. The authors \citep{landman2022classicallyapproximatingvariationalquantum} highlight that in cases of incomplete datasets—where the underlying data distribution is inadequately represented—the PQC may struggle to model it accurately. Expanding on this observation, we emphasize that the surrogate constructed using our proposed method may completely fail if the dataset is incomplete or insufficiently representative. In such cases, the quantum model and its corresponding surrogate could diverge significantly, with little to no similarity in their predictions.

\section{Augmenting dataset beyond available size}\label{sec:appendix_dataset}

Our results (see \cref{fig:training_requirement}) reveal a significant practical challenge: as the number of data points required for surrogation increases with the scale of the quantum model, the size of available datasets becomes a limiting factor. However, this challenge can be addressed using modern machine learning techniques that generate artificial data points with statistical properties similar to the original dataset. These techniques include Variational Autoencoders \citep{augmentation_with_vae}, Generative Adversarial Networks \citep{little2021generativeadversarialnetworkssynthetic}, and the Diffusion Model, which we tested in this context.

To test this model, we utilized a PyTorch-based Diffusion Model from \citep{wang_denoising_diffusion}. The model is a fully connected neural network with an input dimension of 10 and an additional dimension for the time-step embedding. Its core architecture consists of two hidden layers of width \texttt{hidden\_dim} (default 50) with \texttt{ReLU} activations, followed by a final linear layer projecting back to \texttt{input\_dim}.
During training, the time steps $\{t_i\}$ are drawn randomly from $[0, 999]$. At each sampled timestep, the model predicts the noise present in the data, and the mean squared error (\texttt{MSE}) loss function is used to compare the predicted noise with the ground-truth noise. The Adam optimizer is used with a learning rate of $5 \times 10^{-4}$ over $1000$ epochs with a batch size of $16$. After training, new samples are generated through reverse diffusion, allowing the final output to approximate samples from the data distribution learned by the diffusion model.

We used a this Diffusion Model to generate $10,000$ additional artificial data points in addition to the training dataset. This improved the performance of a surrogate that replicates our quantum model trained on a 10 qubit version. By integrating these generated points into the surrogate sampling grid, we observed a substantial improvement in surrogate performance: the relative MSE decreased significantly from $+25\%$ (without artificial data) to approximately $+5\%$. This result highlights diffusion models as a viable technique for enhancing surrogate performance. However, diffusion models come with substantial overhead in terms of extensive hyperparameter tuning and fine-tuning for effective performance. In the future work, it is important to explore which qualities of the dataset (e.g. density, range) play a role in the performance of the surrogation process, which could simplify the process of data augmentation.

\end{appendices}
\end{document}